\documentclass[prl,showpacs,twocolumn]{revtex4}
\usepackage{graphicx}
\begin{document}

\title{Long Range Frustrations in a Spin Glass Model of the Vertex Cover 
  Problem \footnote{The appendix contains an erratum}}

\author{Haijun Zhou}
\affiliation{Max-Planck-Institute of Colloids and Interfaces,
  14424 Potsdam,  Germany}
\affiliation{Institute of Theoretical Physics,
  the Chinese Academy of Sciences, Beijing
  1000190, China
}
\date{\today}
\begin{abstract}
  In a spin glass system on a random graph, 
  some vertices have their spins changing among
  different configurations of a 
  ground--state domain. 
  Long range frustrations may exist 
  among these unfrozen vertices in the sense 
  that certain combinations of spin values for these
  vertices may never appear in any configuration of
  this domain. 
  We present a mean field theory to tackle such long range 
  frustrations and apply it to the NP--hard minimum 
  vertex cover (hard--core gas condensation) problem.  
  Our analytical results on  the ground--state energy
  density and on the fraction of frozen vertices are in
  good agreement with known numerical and mathematical
  results. 
\end{abstract}

\pacs{75.10.Nr, 89.75.-k, 05.20.-y, 02.10.Ox}

\maketitle
\clearpage


The energy landscape of a large spin glass system is very complex.
There may exist (exponentially) many ground--state
and metastable domains in the configurational space; 
these domains are mutually separated from each 
other by (infinitely) high energy barriers.
At low temperature, the system may get  
trapped in one of these configurational
domains, and ergodicity is broken.
In the cavity field formalism \cite{MezardM2001}
of mean field theory of finite connectivity spin glasses
\cite{VianaL1985,KanterI1987,MonassonR1998},
microscopic configurations of a system are 
therefore grouped into different macroscopic
states (hereafter, a macroscopic state is simply referred to as
a $`$state' and a microscopic configuration as a $`$configuration'). 
In a given state each vertex $i$ feels a cavity field 
$h_i$ that may be different for different vertices, and the
fluctuation of this field among  all the states is characterized by a 
probability distribution $P_i(h_i)$ that again 
may be different for different vertices.

The ground--state energy landscape of a spin glass system
can be studied by the zero temperature limit of the cavity field
theory \cite{MezardM2003}. In this
limit and  in a given state $\alpha$, 
the spin value $\sigma_i$ of a vertex $i$ either is
positively frozen ($\sigma_i \equiv +1$ in all configurations)
or is negatively frozen ($\sigma_i\equiv -1$) or is
unfrozen ($\sigma_i$ fluctuates over $\pm 1$ among configurations
of state $\alpha$). A crucial assumption of the cavity field theory
\cite{MezardM2001,MezardM2003} is that, with probability
unity each of the $2^n$ combinations of spin values
for $n$ randomly chosen unfrozen vertices is
realized in configurations of state $\alpha$.
However, this conventional cavity field theory
leads to negative values of structural entropy $\Sigma$
\cite{MezardM2003} when loops of spin--spin interactions
become abundant  (see, e.g., 
\cite{MezardM2002,MezardM2002b,MuletR2002,ZhouH2003b,RivoireO2004b})
or even causes certain type of divergence in the
population dynamics \cite{MezardM2002b,ZhouH2004c}.
To overcome these difficulties,
a positive re-weighting parameter $y$ can be introduced
and its value be determined 
self-consistently by requiring $\Sigma(y)=0$ \cite{MezardM2003}. 
This procedure is however not quite satisfactory; 
in case of the vertex cover problem, 
it predicts a ground--state energy that is systematically
lower than the actual value \cite{ZhouH2003b}.

Here we discuss  the possibility of long range correlations among  
spins of different {\em unfrozen} vertices. 
Both the spins of two unfrozen vertices $i$ and $j$
will certainly fluctuate among configurations of a state $\alpha$. 
On the other hand, we find that with certain probability
$\sigma_i$ and $\sigma_j$ may be 
prohibited to take certain combination of values
({e.g.} $\sigma_i= \sigma_j=-1$) in {\em all} configurations of
state $\alpha$, even if $i$ and $j$ 
are far apart from each other in terms of shortest
path length. 
To detect such long range frustrations among
unfrozen vertices, our idea is to flip the
spin of one unfrozen vertex and then check whether 
this perturbation propagates to other unfrozen ones.
This Letter reports our calculations
on a spin glass model \cite{ZhouH2003b} of the NP-hard minimum vertex cover 
problem \cite{GareyM1979,WeigtM2000,WeigtM2001,HartmannA2003}, which is
equivalent to the hard--core gas condensation of  physics
\cite{BiroliG2002}. 
A long range frustration order parameter 
$R$ is defined.
In this model the quenched randomness comes from the underlying
random graph. Work on systems with additional
quenched randomness of spin--spin interactions is
reported in an accompanying paper \cite{ZhouH2004c}. 

For the vertex cover model, we show 
that long range frustration builds
up ($R>0$) when the mean vertex degree $c$ of the graph exceeds 
$c=e=2.7183$. Analytical predictions on the ground--state 
energy density and on the fraction of frozen vertices are both in 
very good agreement with
known numerical and mathematical results.
The calculations are carried out through the 
cavity approach.  It remains open whether the same 
results are achievable by the replica method.
Our approach is essentially replica symmetric
in the sense that (a) we focus attention on just one 
of all possible macroscopic states, (b) the statistical 
property of this state is specified by just 
three mean field parameters to be defined,
$R$,  $q_+$ and $q_0$. 
Competitions among multiple states
will be included in the theory in future work.

{\bf Some mistakes in the published version of this work are corrected
in the appendix}.


We first introduce the  random graph vertex cover problem.
A random graph $G(N, c)$ has $N$ vertices; 
and between any two vertices an edge is present 
with probability $c/(N-1)$.
The average number of edges incident to a vertex is
$c$ (the mean vertex degree). For large graph size $N$, a vertex's
probability of having $k$ edges is given by the Poisson distribution
$P_{c}(k)=e^{-c} c^k / k !$.
Denote $E(G)$ as the edge set of graph $G$.
A vertex cover of $G$ consists of a set of vertices $\Lambda=\{i_1, 
i_2,\cdots, i_m\}$ such that if edge $(i,j)\in E(G)$, then either 
$i\in \Lambda$ or $j\in \Lambda$ or both. The vertex cover problem consists of
finding a vertex cover $\Lambda$ with size $|\Lambda|\leq n_0$, 
$n_0$ being a prescribed integer. 
This problem is mapped to a spin glass model with energy functional
\begin{equation} 
  \label{eq:energy} 
  E[\{\sigma_i\}]= -\sum{_{i=1}^N} \sigma_i+
  \sum{_{(i,j)\in E(G)}} (1+\sigma_i) (1+\sigma_j) \ .
\end{equation}
$\sigma_i=-1$ if vertex $i\in \Lambda$ (covered) and
$\sigma_i=+1$ otherwise.

The ground--state configurations of model (\ref{eq:energy}) 
correspond to vertex cover patterns with the global minimum 
size \cite{ZhouH2003b}. These configurations may be grouped
into different states \cite{MezardM2003}. 
Two configurations in the same state are mutually reachable by 
flipping a  finite number of spins in one configuration and then 
letting the system relax. (According to this definition of
states, two configurations of the same state
can have a Hamming distance scaling linearly with system size $N$.)
Let us focus on one state, say $\alpha$. In state $\alpha$, 
the spin value of a randomly chosen vertex $i$ may be
fixed to $\sigma_i \equiv +1$, or to $\sigma_i \equiv -1$, or fluctuate
over $\pm 1$.  The fraction of positively frozen, negatively
frozen, and unfrozen vertices in state $\alpha$ is
$q_{+}$, $q_{-}$, and $q_{0}$ respectively.  
(By the way, we notice that
in the minimum vertex cover problem,
the parameters $(q_{+}, q_{-}, q_{0})$ are the same for 
different ground--state states, due to the fact
that the energy density is determined by Eq.~(\ref{eq:minvc}).)
The probability that among $k$ vertices that are
randomly picked up from $G(N, c)$  
$k_0$ are unfrozen, $k_+$ positively frozen, 
and $k_-$ ($=k-k_0-k_+$) negatively
frozen is $k!/(k_{0}! k_{+}! k_{-}!) q_0^{k_0} q_{+}^{k_+}
q_{-}^{k_-}$ (in the large $N$ limit).

Since the spin of an unfrozen vertex $i$ fluctuates among
different configurations of state $\alpha$, the $`$correlation length' 
of this fluctuation is an important issue.
We ask the following question: If $\sigma_i$ is externally
fixed to $\sigma_i=-1$, how many other unfrozen vertices must
eventually fix their spins as a consequence? 

For a random graph of size $N\to \infty$, 
the total number $s$ of affected vertices may scale linearly with $N$.
If this happens, vertex $i$ is referred to as
type-I unfrozen. The probability for this to 
happen is denoted as $R$ (which defines 
our long range frustration order parameter). 
The total number of affected vertices may also be finite. 
In this case, vertex $i$ is  type-II unfrozen.  Based on insights gained from
studies on random graphs \cite{BollobasB1985}, we know that the
percolation clusters evoked by two type-I unfrozen vertices 
have a nonzero intersection (of size proportional to $N$).
Therefore, the spin values of all the type-I unfrozen vertices
must be strongly correlated. If we randomly choose two type-I
unfrozen vertices $i$ and $j$, then with probability {\em one-half}
their spin values can not be negative simultaneously: if 
$ \sigma_i = -1$, then $\sigma_j$ must be $+1$; 
if $\sigma_j = -1$, then $\sigma_i$ must be $+1$.
On the other hand, two randomly chosen type-II unfrozen 
vertices are mutually independent, since each vertex can only 
influence the spin values of $s \sim O(1)$ other unfrozen vertices
while the shortest path length between two randomly chosen vertices 
of $G(N, c)$ scales as $\ln N$ and becomes divergent when 
$N \to \infty$ \cite{BollobasB1985}. 
Denote $f(s)$ as the probability that a randomly chosen unfrozen
vertex $i$, when flipped to $\sigma_i = -1$, 
will eventually fix the spin values of $s$ 
unfrozen vertices with $s$ being finite
and therefore $\lim_{N\to \infty} s/N=0$.

We calculate the parameters $q_0$, $q_{+}$, $q_{-}$ by the
cavity field method \cite{MezardM2001,MezardM2003}:
First a random graph $G(N, c)$ is generated; then a new vertex $i$
is connected to a set $V_i$ of $k$ randomly chosen vertices 
in $G(N, c)$, $k$ following the distribution $P_c(k)$; the 
unfrozen/frozen probabilities $\{q_0(i), q_{+}(i), q_{-}(i)\}$ 
of vertex $i$ in the
enlarged graph (denoted as $G^\prime$) is then calculated.
We assume  the following convergence condition: 
$\lim_{N\to \infty} \{ q_0(i), q_{+}(i), q_{-}(i)\}=\{q_0, q_{+}, q_{-}\}$.
This enables us to write down
a set of self-consistent equations in the large $N$ limit.

If the new vertex $i$ is positively frozen, then 
none of the vertices in $V_i$ are positively frozen in graph $G$.
Furthermore, there are two possible situations:
{\bf (i)}  no vertices in $V_i$ is type-I unfrozen in $G$;
or {\bf (ii)}  some of the vertices in $V_i$  are type-I unfrozen.
In case (ii), all these type-I unfrozen vertices will take
spin value $-1$ simultaneously in some configurations of state $\alpha$, so that
vertex $i$ will have $\sigma_i \equiv +1$ as it is added into the system. With
this analysis, we get a self-consistent equation for $q_{+}$:
\begin{eqnarray}
  q_{+} &=& \sum\limits_{k=1}^{\infty}  P_c(k) \sum\limits_{l=1}^{k} C_k^l
  \bigl( q_{0} R \bigr)^l
  \bigl( q_{0} ( 1 - R ) + q_{-} \bigr)^{k - l} 2^{1-l} \nonumber \\
  & & + \sum\limits_{k=0}^{\infty} P_c(k) \bigl( q_{0}( 1 - R) + q_{-} \bigr)^{k}
  \nonumber \\
  &=&   2 e^{ - c  q_{+}  - (1/2) c q_0 R } 
  - e^{ -c q_{+} - c q_0 R } \ ,
  \label{eq:q+}
\end{eqnarray}
where $C_k^l = k!/(l! (k-l)!)$. Equations (\ref{eq:q+}),
(\ref{eq:q0}), (\ref{eq:fs}) and (\ref{eq:backbone})
 are derived elsewhere \cite{supplementary}.

If the new vertex $i$ is unfrozen, there are also two possibilities 
concerning the spin values of vertices in $V_{i}$:
{\bf (iii)} none of them is positively frozen in $G$;
or {\bf (iv)} one of them is positively frozen in $G$. 
To ensure vertex $i$ will be unfrozen,
in situation (iii) two or more of the vertices in $V_{i}$ must be
type-I unfrozen in $G$, among which one is in conflict 
with all the others; and 
in situation (iv) some of the vertices in $V_i$ may be type-I unfrozen 
in $G$, but they must be capable of taking spin value $-1$ simultaneously. 
Therefore, we get a self-consistent equation for $q_0$
\cite{supplementary}:
\begin{eqnarray}
  \label{eq:q0}
  q_0 &=&  \bigl( 2 c q_{+} + c q_0 R \bigr)   e^{ - c q_{+} - (1/2) c q_{0} R } 
  \nonumber \\
  & &  -\bigl( c q_{+} + c q_0 R +  ( c q_0 R )^2/4
  \bigr) e^{ - c q_{+} -c q_0 R} \ .
\end{eqnarray}

If the spin of an unfrozen vertex $i$ is flipped to $\sigma_i=-1$, 
it may not affect any other vertices ($s=0$), provided its local
environment is described by the above mentioned situation (iii). 
This happens with probability $p_1 = 1 - c q_{+}^2 / q_0$.
With probability $1-p_1$, the unfrozen vertex $i$ encounters a local
environment of type (iv), that is, one of
its nearest neighbors vertex $j$ is positively frozen in graph $G$.
This vertex $j$ must face the local environment of 
type (i) in graph $G$ if vertex $i$ is type-II unfrozen. 
(If vertex $j$ has the local environment of type (ii), flipping
the spin value of vertex $i$ to $\sigma_i=-1$ would cause
a percolation cluster of size proportional to $N$.) 
With these preparations, we obtain the following 
self-consistent equation for the distribution $f(s)$
\cite{supplementary}:
\begin{equation}
  \label{eq:fs}
  f(s)= p_1 \delta_{s}^{0} + (1 - p_1 )
  \sum\limits_{l=0} P_{c^{\prime}}( l )
  \sum\limits_{\{s_m\}} \prod\limits_{m}^{l} f(s_m) \delta_{s-1}^{s_1+\ldots+s_l} \ .
\end{equation}
In Eq.~(\ref{eq:fs}), $\delta$ is the Kronecker symbol;
and $c^\prime= c q_0( 1 - R )$ is 
the mean number of type-II unfrozen vertices adjacent
to a positively frozen vertex.
Since $R=1-\sum_{s=0}^{\infty} f(s)$, we establish that 
the long range frustration order parameter $R$ is determined by the following 
equation:
\begin{equation}
  \label{eq:R}
  R= ( c q_{+}^2 / q_0 )  \bigl( 1- e^{- c q_0 R (1-R)}  \bigl) \ .
\end{equation}
A positive $R$ signifies the appearance of a percolation cluster
of unfrozen vertices whose spin values are strongly correlated.

\begin{figure}[b]
\vskip -0.7cm
\includegraphics[width=0.8\linewidth,angle=270]{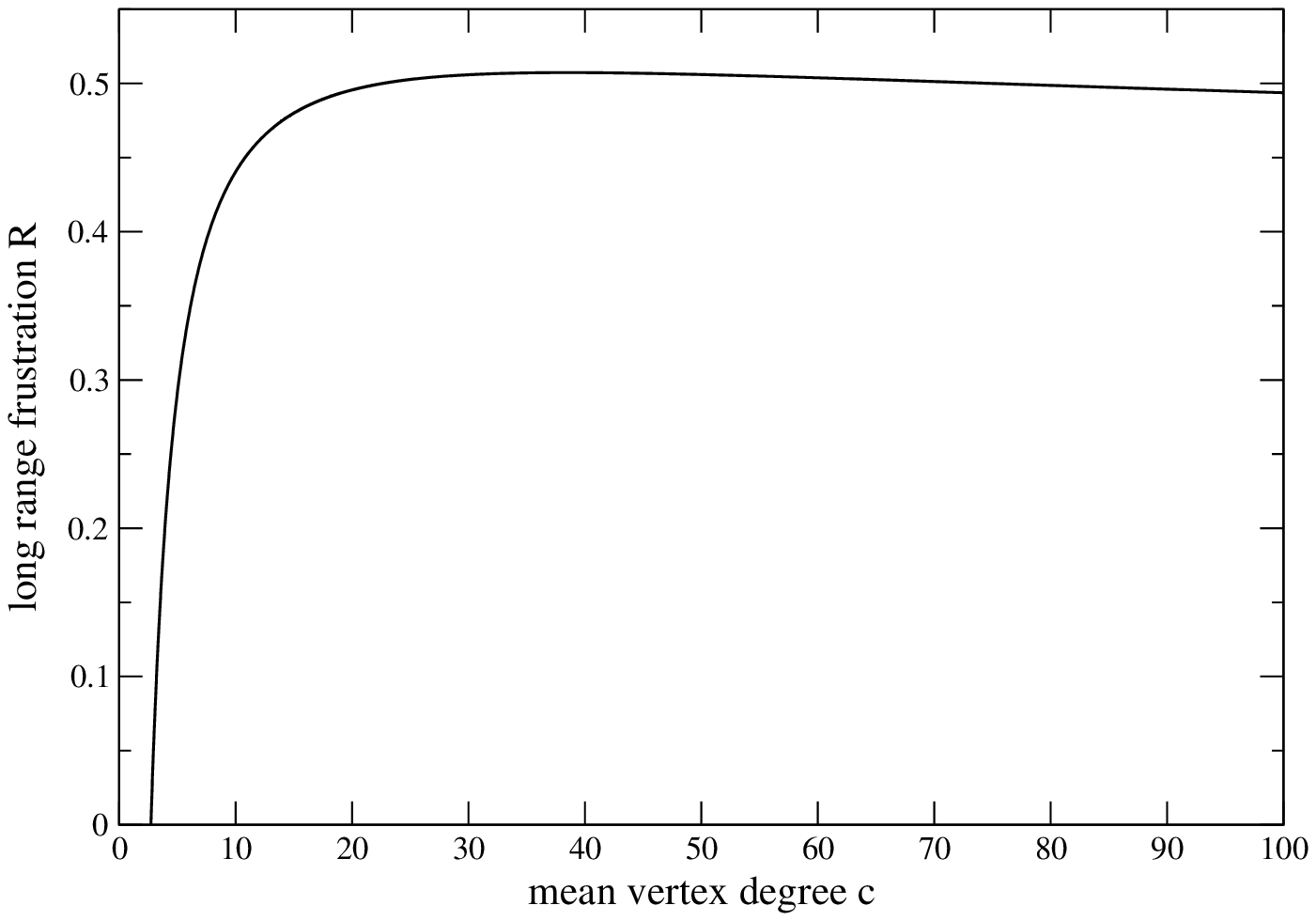}
\caption{
  \label{fig:R}
  The long range frustration order parameter $R$.
}
\end{figure}

Figure \ref{fig:R} shows the value of $R$
as a function of  mean vertex degree $c$. $R\equiv 0$ when $c\leq e$;
this is consistent with Ref.~\cite{BauerM2001} that,
a minimal vertex cover pattern can be found by a polynomial 
leaf--removal algorithm.  When $c > e$, a finite fraction of the
unfrozen vertices are long--rangely frustrated; the leaf--removal
algorithm outputs a looped subgraph \cite{BauerM2001}.
At mean vertex degree $c\simeq 40$, 
the order parameter $R$ reaches a maximal value;
then it gradually decays as $c$ is 
further increased. 

\begin{figure}[b]
\vskip -0.7cm
\includegraphics[width=0.8\linewidth,angle=270]{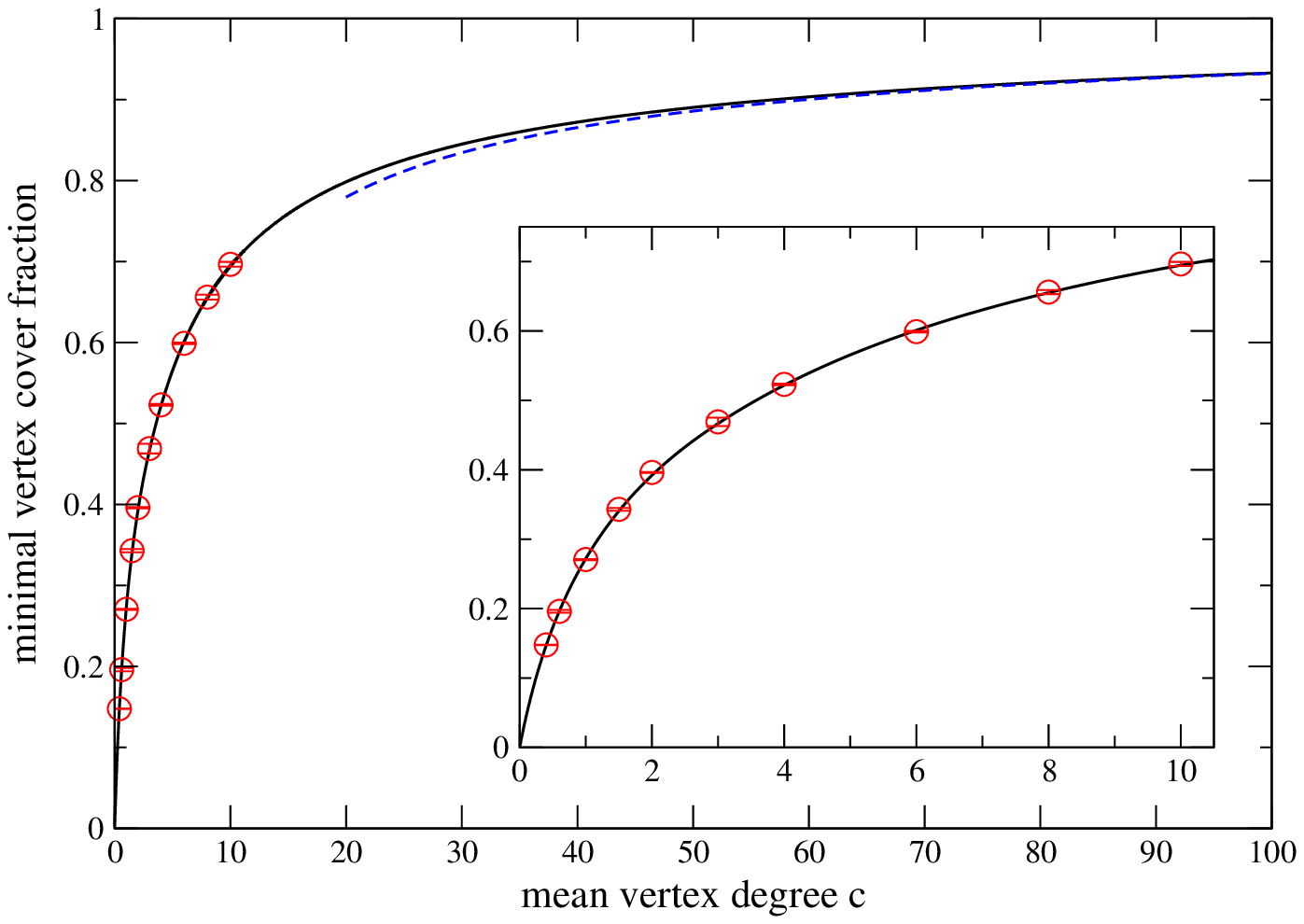}
  \caption{
    \label{fig:minvc}
    The minimal vertex cover fraction $X_{\rm min}$ (Eq.~(\ref{eq:minvc}), 
    solid line) and its comparison
    with the asymptotic formula of Ref.~\cite{FriezeA1990}
    (dashed line) and the numerical results of Refs.~\cite{WeigtM2000} 
    (symbols).
  }
\end{figure}

The fraction of vertices that are covered in a minimal vertex cover
is \cite{supplementary}
\begin{equation}
  \label{eq:minvc}
  X_{\rm min}=1-q_{+}- {q_{0} / 2} \ .
\end{equation}
Figure \ref{fig:minvc} shows the relationship between $X_{\rm min}$
and mean vertex degree $c$. At large $c$ values,
Eq.~(\ref{eq:minvc}) is in agreement with a rigorous
asymptotic expression given by Frieze \cite{FriezeA1990}; at low
values of $c$, it is in
agreement with the exact   enumeration results of
Weigt and Hartmann \cite{WeigtM2000}. 
These excellent agreements are quite encouraging, 
in view of the fact that all previous efforts failed
\cite{WeigtM2000,WeigtM2001,ZhouH2003b}.
It has already been
established that  when $c > e$ the replica symmetric
solution of vertex cover problem becomes unstable 
\cite{WeigtM2000,WeigtM2001}; but earlier 
replica symmetry breaking
solutions either resulted in
negative structural entropy or predicted a minimal vertex cover
size noticeably lower than the actual value \cite{ZhouH2003b}.

\begin{figure}[t]
  \includegraphics[width=0.8\linewidth,angle=270]{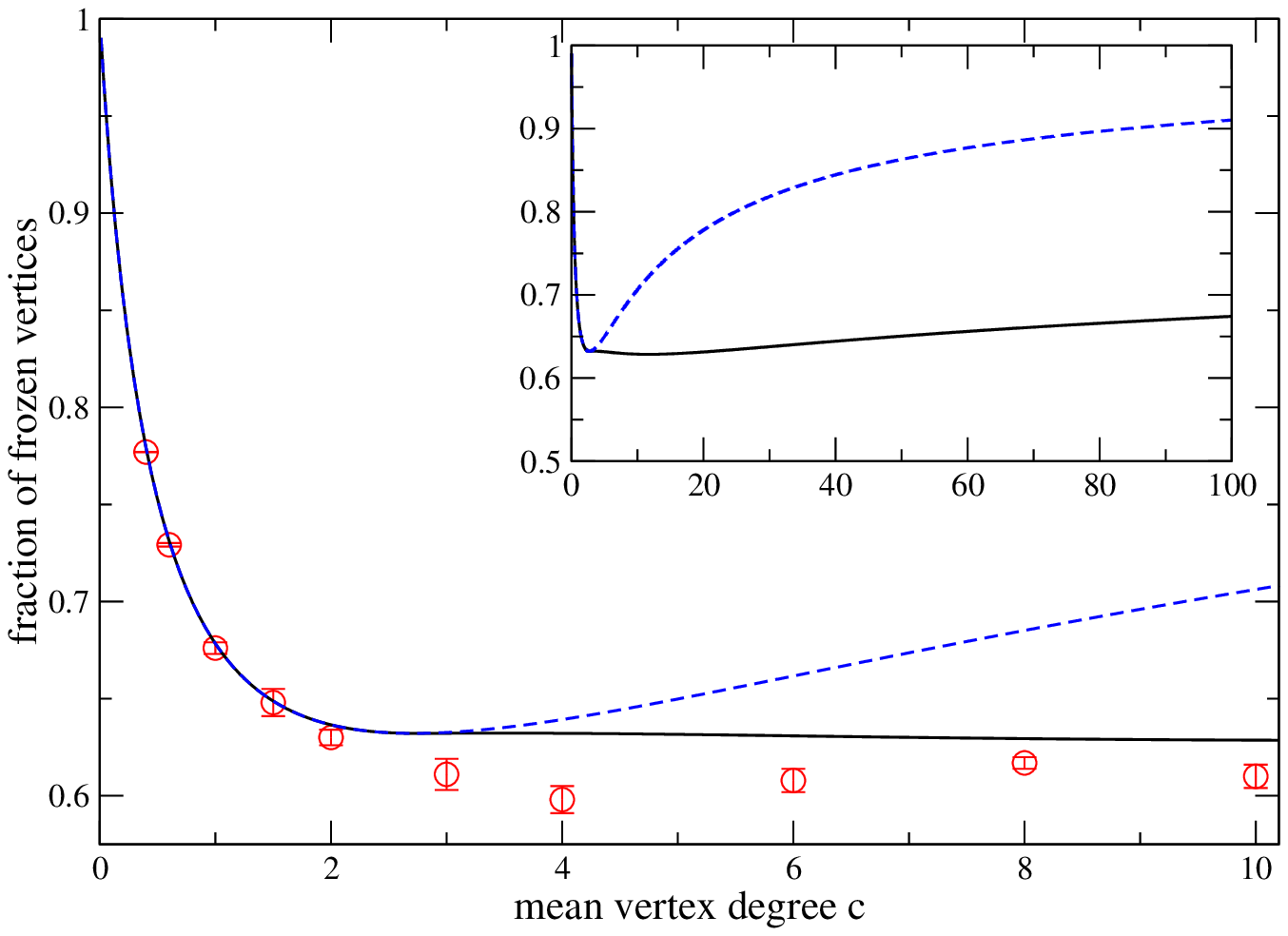}
  \caption{
    \label{fig:backbone}
    Fraction of frozen vertices in {\em all} 
    macroscopic states 
    (Eq.~(\ref{eq:backbone}), solid lines)  and its
    comparison with the numerical results of Ref.~\cite{WeigtM2001}
    (symbols) and the fraction of frozen 
    vertices in  {\em one} macroscopic state (dashed lines).
  }
\vskip -0.5cm
\end{figure}

So far we have focused on only one ground--state  state
of the vertex cover problem. When $c > e$ it is believed that
there are many such states (replica symmetry breaking).
This is consistent with our observation that, when $c > e$ 
the fraction of frozen vertices ($=q_{+}+q_{-}$,
dashed lines in Fig.~\ref{fig:backbone}) in one state
is much higher than the actual fraction of frozen vertices
estimated numerically 
(symbols in Fig.~\ref{fig:backbone}) 
\cite{WeigtM2001}.
This is easy to understand: A frozen vertex in
one state may be unfrozen or be frozen to the opposite spin value in
another state. At the moment we are unable to
construct a theory to include the competitions among
different states.
As a first attempt, we make the following conjectures: 
(a) if a vertex is positively frozen in one state, it is positively
frozen in all states; and (b) a vertex is negatively frozen
in all states only if it is adjacent to two or more positively
frozen vertices. 
Then an expression
on the fraction  of frozen vertices is obtained
\cite{supplementary}:
\begin{equation}
  \label{eq:backbone}
  \Gamma = q_{+} + 1 - e^{- c q_{+}} - c q_{+} e^{- c q_{+}} \ .
\end{equation}
The agreement of Eq.~(\ref{eq:backbone}) with  the numerical 
data of Ref.~\cite{WeigtM2001} is quite good (Fig.~\ref{fig:backbone}). 
This is an issue to be understand more deeply.

To summarize, we have studied 
long range frustrations among {\em unfrozen} vertices 
in a macroscopic state of a spin glass system.
We found that, with certain probability,
the fluctuations of the spin values of two
or more distantly separated unfrozen vertices 
are highly correlated. 
A long range frustration order parameter $R$ was
calculated to quantify this strong correlation.
When applying our method to the
NP--hard minimum vertex cover (hard--core
gas condensation) problem, the  analytical predictions
concerning the ground--state energy density 
and the fraction of frozen
vertices are in good agreement with known numerical and rigorous 
results. The basic idea behind this paper is also applicable
to other spin glass systems \cite{ZhouH2004c}.

We emphasize that the appearance of many macroscopic states 
in the energy landscape of a spin glass system does not necessarily mean 
the existence of long range frustrations among
unfrozen vertices in a single macroscopic state.
As an counter--example, in the maximum matching problem 
\cite{ZhouH2003d} there is no long range frustrations ($R\equiv 0$) 
but there exist an exponential number of macroscopic states. 
It is interesting to notice that the maximum matching
problem can be solved by polynomial algorithms.  
It appears that, the proliferation of macroscopic states is not the real 
reason of the computational complexity in finding a ground--state 
configuration for a disordered system.
As another example, there are many macroscopic states in a
typical random $3$-satisfiability formula when
$3.921 < \alpha < 4.267$ (here $\alpha$ is the
clauses-to-variables ratio); but the
survey propagation algorithm is able to find a solution efficiently
\cite{MezardM2002,MezardM2002b}. 

On the other hand, we believe the existence of long range frustrations among
unfrozen vertices will make it intrinsically difficult for a search algorithm
to find a ground--state configuration. Because of these
long range effects, it is difficult (a) to determine whether a
vertex is frozen or unfrozen in a macroscopic state and, (b)
to trace the percolation cluster associated with a given
unfrozen vertex. Recently, some NP-hard 
combinatorial optimization 
problem in computer science were
studied by zero temperature cavity field method
\cite{MezardM2002,MezardM2002b,MuletR2002,ZhouH2003b}. 
We hope the present work,
besides improving our understanding of finite connectivity spin glasses,
will stimulate further efforts in finding more efficient
algorithms. We are presently implementing the physical
picture of this paper into an algorithm for the vertex cover problem.


{H.Z.} is grateful to R.~Lipowsky,
Z.-C.~Ou-Yang, and L.~Yu for their support,
to A.~K.~Hartmann and M.~Weigt for sharing
their numerical results, and to L.~Yu and
R.~Lipowsky for helpful discussions.


\vskip 4.5cm
\section*{Erratum: Long-Range Frustration in a Spin-Glass Model of the Vertex-Cover Problem [Phys. Rev. Lett. 94, 217203 (2005)]
}

We found several mistakes in the Letter \cite{Zhou-2005a}, which
significantly affect the quantitative theoretical results.

\begin{figure}[b]
\vskip 1.0cm
\begin{center}
\includegraphics[width=0.8\linewidth]{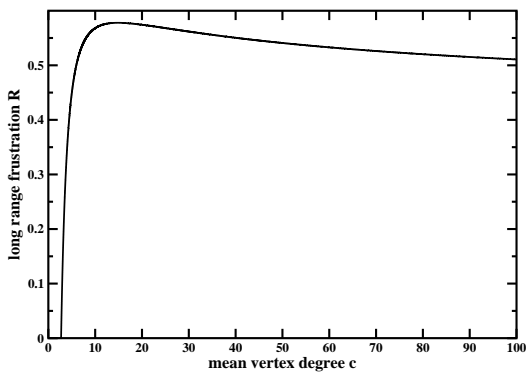}
\end{center}
\caption{
\label{fig:Rvalue}
(Corrected version of Fig.~1) The long-range frustration order parameter $R$. 
}
\end{figure}

The expression (5) for the long-range frustration order parameter $R$
was wrong. For an unfrozen vertex $i$ to be type-I unfrozen 
in graph $G$, one of its nearest neighbors (say $j$) must be positively frozen
and facing the local environment of type (ii) in $G^\prime$ (the
graph obtained by removing $i$ and its edges from $G$).
The correct formula for $R$ is
$$
R = \frac{c q_{+}^2}{q_0}
\Bigl(1- \frac{1}{q_{+}}
e^{- c q_{+} - c q_0 R} \Bigr) .
$$
$R$ as a function of the mean vertex
degree $c$ is shown in the corrected figure (Fig.~\ref{fig:Rvalue}).
It is positive for $c > e = 2.718\ldots$ and
its maximum is reached at $c \simeq 14.85$.

The self-consistent equation (4) for the size distribution $f(s)$
was  wrong. The
quantity $p_1$ of this expression should be replaced by
$p_1^\prime = p_1/(1-R)$, which is the
conditional probability that a vertex $i$ faces
the environment of type (iii) given that 
$i$ is a type-II unfrozen vertex in $G$. The function $f(s)$
satisfies $\sum_{s=0}^{\infty} f(s) = 1$.

\begin{figure}
\begin{center}
\includegraphics[width=0.8\linewidth]{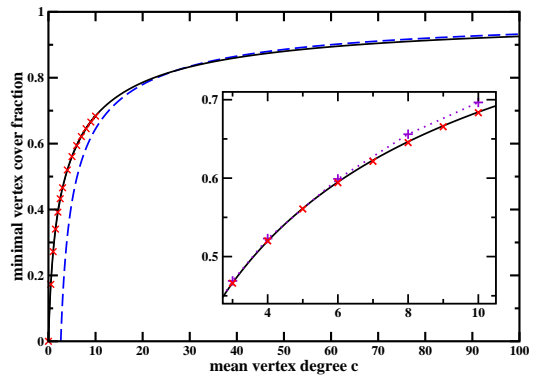}
\end{center}
\caption{
\label{fig:Xvalue}
(Corrected version of Fig.~2). The minimal vertex-cover fraction
$X_{\rm min}$ (solid line) and
its comparison with the
asymptotic formula of Ref.~\cite{Frieze-1990} (dashed line),
the numerical results of Ref.~\cite{Weigt-Hartmann-2000} (`+' symbols) 
and  Ref.~\cite{Weigt-Zhou-2006} (`$\times$' symbols).
}
\end{figure}

Equation (6) for the fraction of
covered vertices $X_{\rm min}$ was also
wrong. According to the analysis in
\cite{Zhou-2005b}, the correct formula should be
$$
X_{min} = \frac{1}{c} \int\limits_{0}^{c} \bigl[1- q_{+}(c^\prime)
\bigr] {\rm d} c^{\prime} ,
$$
where $q_{+}(c)$ is the fraction of positively frozen
vertices at mean vertex degree $c$. 
$X_{\rm min}$ as a function of $c$ is shown in the
updated figure (Fig.~\ref{fig:Xvalue}). The theoretical prediction
is in agreement with simulation results obtained on
single large graphs \cite{Weigt-Zhou-2006}, it is slightly
lower than the enumeration result obtained on single small graphs 
\cite{Weigt-Hartmann-2000}. 
When the mean vertex degree $c$ is large, the value
of $X_{\rm min}$ is slightly lower than the
asymptotic result of \cite{Frieze-1990}.

Financial support from the Chinese Academy of Sciences
(KJCX2-EW-J02) and NSFC (grant numbers 10834014 and 11121403) is
acknowledged.


\end{document}